\documentclass[12pt,preprint]{aastex}
\newcommand{\etal}{\mbox{\rm et al.~}}
\newcommand{\ms}{\mbox{m s$^{-1}~$}}
\newcommand{\kms}{\mbox{km s$^{-1}~$}}

\newcommand{\mse}{\mbox{m s$^{-1}$}}

\newcommand{\msun}{M$_{\odot}~$}
\newcommand{\msune}{M$_{\odot}$}

\newcommand{\lsun}{L$_{\odot}~$}
\newcommand{\mjup}{M$_{\rm JUP}~$}
\newcommand{\mjupe}{M$_{\rm JUP}$}
\newcommand{\msat}{M$_{\rm SAT}~$}
\newcommand{\msate}{M$_{\rm SAT}$}
\newcommand{\mnep}{M$_{\rm NEP}~$}
\newcommand{\mnepe}{M$_{\rm NEP}$}

\newcommand{\mearthe}{M$_{\rm Earth}$}

\newcommand{\rjupe}{R$_{\rm JUP}$}
\newcommand{\msini}{$M \sin i~$}
\newcommand{\vsini}{$v \sin i~$}

\newcommand{\chisq}{$\chi_{\nu}^2$}
\newcommand{\arel}{$a_{\rm rel}$}

\newcommand{\rphk}{\ensuremath{R'_{\mbox{\scriptsize HK}}}}
\newcommand{\lrphk}{\ensuremath{\log{\rphk}}}
\newcommand{\caii}{\ion{Ca}{2} H \& K}
\newcommand{\mv}{\ensuremath{M_{\mbox{\scriptsize V}}}}

\received{}
\accepted{}

\slugcomment{ }

\shortauthors{Butler {\it et~al.\/}}
\shorttitle{A Neptune--Mass Planet}
\begin{document}

\title{A Neptune--Mass Planet Orbiting \\ the Nearby M Dwarf GJ~436$^{1}$}
\author{R. Paul Butler\altaffilmark{2},
Steven S. Vogt\altaffilmark{3}, 
Geoffrey W. Marcy\altaffilmark{4}, 
Debra A. Fischer\altaffilmark{4,5}, 
Jason T. Wright\altaffilmark{4}
Gregory W. Henry\altaffilmark{6},
Greg Laughlin\altaffilmark{3}
Jack J. Lissauer\altaffilmark{7}}

\email{paul@dtm.ciw.edu}

\altaffiltext{1}{Based on observations obtained at 
    the W.M. Keck Observatory, which is operated jointly
    by the University of California and the California
    Institute of Technology.  Keck time has been granted by
    both NASA and the University of California.}

\altaffiltext{2}{Department of Terrestrial Magnetism, Carnegie Institution of
Washington, 5241 Broad Branch Rd NW,  Washington DC, USA 20015-1305}

\altaffiltext{3}{UCO/Lick Observatory, 
University of California at Santa Cruz, Santa Cruz, CA, USA 95064}

\altaffiltext{4}{Department of Astronomy, University of California,
Berkeley, CA USA  94720}

\altaffiltext{5}{Department of Physcis and Astronomy, San Francisco State University,
San Francisco, CA, USA 94132}

\altaffiltext{6}{Center of Excellence in Information Systems,
Tennessee State University, 330 10th Avenue North, Nashville, TN  37203;
Also Senior Research Associate, Department of Physics and Astronomy, 
Vanderbilt University, Nashville, TN  37235}

\altaffiltext{7}{Space Science Division, 245-3 NASA Ames Research
  Center, Moffett Field, CA 94035-1000 USA}

\begin{abstract}
We report precise Doppler measurements of GJ~436 (M2.5V) obtained at
Keck Observatory.  The velocities reveal a planetary companion with
orbital period of 2.644 d, eccentricity of 0.12 (consistent with zero)
and velocity semi-amplitude of $K =18.1$ \ms.  The minimum mass (\msini)
for the planet is 0.067 \mjup = 1.2 M$_{\rm NEP}$ = 21 M$_{\rm EARTH}$,
making it the lowest mass exoplanet yet found around a main sequence
star and the first candidate in the Neptune mass domain.  GJ~436 (Mass
= 0.41 \msune) is only the second M dwarf found to harbor a planet,
joining the two--planet system around GJ 876.  The low mass of the
planet raises questions about its constitution, with possible
compositions of primarily H and He gas, ice/rock, or rock--dominated.
The implied semi--major axis is $a$ = 0.028 AU = 14 stellar radii,
raising issues of planet formation, migration, and tidal coupling with
the star.  GJ~436 is $>3$ Gyr old, based on both kinematic and
chromospheric diagnostics. The star exhibits no photometric
variability on the 2.644-day Doppler period to a limiting amplitude of
0.0004 mag, supporting the planetary interpretation of the Doppler
periodicity.  Photometric transits of the planet across the star are
ruled out for gas giant compositions and are also unlikely for solid
compositions.  As the third closest known planetary system, GJ~436
warrants follow--up observations by high resolution optical and IR
imaging and by the Space Interferometry Mission.

\end{abstract}


\section{Introduction}
\label{intro}

To date, $\sim$135 extrasolar planets are securely known around nearby
FGKM stars.  All were discovered by the Doppler technique (see
references within Butler \etal (2002) and Mayor and Santos
(2003).\footnote{References to published papers and updates on orbital
parameters can be found at \url{http://exoplanets.org/}}).  The minimum
masses span the range from 0.1 \mjup to above 13 \mjupe, merging into
the ``brown dwarf'' domain.  The distribution of planet masses rises
steeply toward the lowest detectable masses with a power law
dependence, d$N$/d$M \propto M^{-1.3}$ \citep{Marcy_Butler00,
Marcy04a}, even after correction for the unknown orbital inclination
(Jorissen, Mayor \& Udry 2001).  Two of the domains that remain relatively
unexplored are the distribution of planet masses below 1
\msat and the occurrence of planets in general for low mass stars.

The distribution of masses of planets below that of Saturn remains poorly
constrained because of the difficulty in their detection, demanding
Doppler precision of 3 \ms or better.  Prior to the discovery reported
herein, eight known exoplanets had \msini
below 1 \msate, namely those orbiting HD~16141 and HD~46375 (Marcy
\etal 2000), HD~16874 (Pepe \etal 2002), HD~76700 (Tinney \etal 2003),
HD 49674 (Butler \etal 2003), HD 3651 (Fischer et al. 2003), 
55 Cnc (planet ``c'', Marcy et al. 2002) and HD 99492 \citep{Marcy04b}.  HD
46375 and HD 99492 have the lowest known minimum masses, both having
\msini = 0.11 \mjup = 0.3 \msat.  Two sub--saturn candidates reside in
distinctly eccentric orbits, namely HD 16141 ($P$=75 days, $e$=0.18)
and HD 3651 ($P$=62.2 d, $e$ = 0.64), suggesting that whatever
mechanism pumps eccentricities in exoplanets, it acts on planets of
sub--Saturn mass as well as on planets of $\sim$10 \mjup.

Only one planetary system was previously known around an M dwarf, 
GJ~876 with its two planets (Marcy et al. 2001).  The total
number of M dwarfs being surveyed by precise Doppler measurements is
roughly 200 (Wright et al. 2004, Mayor \& Santos 2003, Kurster et
al. 2003, Endl et al. 2003).  The solitary planetary system (GJ~876)
known around M dwarfs implies that the occurrence rate of planets
having masses greater than 1 \mjup and orbiting with periods, $P <$ 3
yr ($a<$ 1.5 AU), is only $\sim$0.5 \%.  In contrast, the occurrence
rate of analogous planets around F \& G--type main sequence stars is $\sim$5\%
(Mayor \& Santos 2003; Marcy et al. 2004a) for such orbital periods.
Thus, the occurrence rate of Jupiter--mass planets around M dwarfs
having masses, $M$=0.3--0.5 \msun is roughly an order of magnitude
lower than that around F \& G main sequence stars with $M$ = 0.8--1.2
\msun.

Obviously, this approximate estimate of the occurrence of jupiters as
a function of stellar mass suffers both from selection effects and
from small numbers.  The faintness of M dwarfs makes Doppler
measurements more difficult.  Nonetheless, planets of Jupiter--mass
would make a Doppler signature of at least 20 \ms for orbits within 2
AU, rendering them easily detectable.  Thus, the decline in planet
occurrence with smaller stellar mass has statistical
integrity, although the data are insufficient to establish an accurate
relationship.

The diversity of planetary systems, including their observed masses and
orbits, almost certainly stems from formation processes
in protoplanetary disks (see for example, Lissauer 1995; 
Levison, Lissauer, \& Duncan, 1998; Alibert, Mordasini, \& Benz 2004).
Young, low mass stars may be surrounded by protoplanetary disks that have
lower mass and lower surface mass density than those surrounding young
solar mass stars (T Tauri stars).  If so, the formation of
Jupiter--mass planets may be inhibited at all orbital radii (Laughlin,
Bodenheimer, Adams 2004).  Thus, low mass planets of Neptune--Saturn mass
may be as common around M dwarfs as Jupiter--mass planets are around
solar mass stars.  Alternatively, gas accretion may operate so much
less efficiently, or the lifetime of the gas may be so short, in low
mass protoplanetary disks that predominantly rock--ice cores form
only, with few gas giants forming.

More M dwarfs should be surveyed to establish the dependences of
planet properties on stellar mass, especially at detection thresholds
of sub--Saturn masses.  Here, we report the detection of a planet with
the lowest \msini yet found, orbiting an M dwarf.

\section{Properties of GJ~436}
\subsection{Mass and Age}

GJ~436 (HIP 57087, LHS 310) is an M2.5V star with V = 10.67, B--V = 1.52
and a parallax of 97.73 mas ($d$ = 10.23 pc) with an uncertainty of
3\% from Hipparcos (ESA 1997) implying an absolute visual magnitude,
\mv = 10.63, consistent with typical field M dwarfs of its spectral
type residing on the main sequence.

Its mass may be estimated from various empirical mass--luminosity
calibrations and theoretical models for M dwarfs.  The empirical
relation between stellar mass and \mv\ from Henry \& McCarthy (1993)
suggests that $M_{\star}$ = 0.42 $\pm$ 0.05 \msune, where the uncertainty stems
from the standard deviation of the observed scatter in the stars of
measured mass at a given \mv.  Benedict et al. (2001) provide an
updated mass--luminosity relation which agrees well with that from
Henry \& McCarthy.  Delfosse et al. (2000) use their newly discovered
M dwarf binaries, the calibration from which yields a mass for GJ~436 
of 0.40 $\pm$ 0.05 \msune, in good agreement with that
of Henry \& McCarthy.  Theoretical models from Baraffe et al. (1998)
for solar metallicity, [M/H]~=~0.0, predict a mass of 0.40 \msun for GJ~436.  
Their models suggest that if the star were metal--rich,
[M/H]~=~+0.25, the implied mass would be higher by approximately 0.05
\msun, constituting a plausible systematic error due to the unknown
metallicity of the star.  The models of Siess et al. (2000) suggest a
mass of 0.35 \msune, somewhat less than that from the empirical
relation.  Here we adopt the simple average of the two empirical
estimates from Henry \& McCarthy (1993) and from Delfosse et
al. (2000) for GJ~436 yielding $M$=0.41 $\pm$ 0.05 \msun.

The age of GJ~436 may be constrained by various diagnostics.
Leggett (1992) reports Galactic UVW velocities of +45.3, -20.0, +17.9
\kms, in agreement with Reid, Hawley, \& Gizis (1995) who give +44,
-20, +20.  These velocities render the star a member of the ``old
disk'' population.  Indeed, for M dwarfs fainter than V = 10, there
remains a kinematic bias toward identification of older, metal--poor
stars and subsequent inclusion in catalogs (Reid, Hawley, Gizis 1995;
Carney, Latham \& Laird 1990).  Its UVW velocity components suggest
that GJ~436 has an age of at least 2 Gyr and probably has a metal
abundance not much greater than solar.  Furthermore, GJ~436 is not
a flare star nor does it exhibit particularly strong chromospheric
emission at \caii ~(Figure 1) for an M dwarf of its spectral type.
It shows no emission at the Balmer lines in our high resolution
spectra.  This low chromospheric activity is consistent with a star
middle-aged or older, placing the age likely greater than 3 Gyr
consistent with the kinematics.  We measured the rotational line
broadening to be, \vsini $<$ 3 \kms (\S 5).  Also, GJ~436 is
photometrically constant at millimag levels (see \S 6), indicating
that spots and magnetic fields are weak, consistent with an age
greater than 3 Gyr.

\subsection{Velocity Jitter and Ca II H\&K }

The photospheric velocity jitter of GJ~436 may be estimated from the
$\sim$30 other M dwarfs on our Keck planet--search program that have
similar stellar properties, namely B-V between 1.4 and 1.6, similar
\mv, and similar \caii \ emission, as described by Wright et al.
(2004), and have been observed at least 10 times over 4 or more
years.  In brief, the $\sim$ 30 M dwarfs of similar spectral type,
\mv, and \rphk \ are deemed comparison stars. For each of them the RMS
of their velocity measurements is determined and the internal velocity
error is subtracted in quadrature.  The remaining velocity scatter
represents the variance of our velocities caused by all sources
excluding photon--limited errors.  A minority of our comparison M
dwarfs may have unseen companions which would raise the velocity RMS
above that caused simply by photospheric jitter.  Thus, Wright et
al. determine the median value (instead of the mean) of those RMS
values for the comparison stars, to suppress the effect of companions.
The resulting median of the RMS values is found to be 3.3 \mse, with a
standard deviation of 2.1 \ms about that median.  This estimated
jitter of 3.3 \ms is presumably due to convective overshoot, spots,
flares, oscillations, and other non--uniformities on the rotating
stellar surface.  However, this jitter estimate, by its empirical
construction, also includes errors caused by any instrumental and
software inadequacies, as well as by low mass planets, that cause
variations in our actual velocity measurements.

Endl et al. (2003) obtained 17 radial velocity measurements of
GJ~436 using the Hobby--Eberly Telescope during 394 days.
Those velocities exhibited an RMS of 20.6 \mse, constituting an excess
velocity variability above errors.  Their search for periodicities did
not a reveal significant signal.  Endl et al. (2003) considered
carefully the possibility that jitter was the cause of the excess
velocity scatter.  They note that its stellar magnetic activity level
and X--ray luminosity ($L_{\rm X} = 0.6 \times 10^{27} {\rm ergs-s}^{-1}$)
are modest and similar to other quiet, old M dwarfs (such as GJ 411)
that show a velocity RMS of less than 10 \ms, suggesting that the jitter
of GJ~436 should be similarly smaller than 10 \ms.  Our analysis
similarly finds that GJ~436 is only modestly active, consistent
with jitter of $\sim$ 3 \mse.  Thus, it is quite possible that some of the
excess velocity scatter noted by Endl et al. (2003) was caused by the
planet we detect here.  Our velocities (presented in Table 1) are not
inconsistent with the results plotted for GL 436 in Figure 6 of Endl
et al. within their error bars of 12--19 \mse.

The chromospheric emission at the {\ion{Ca}{2} K} line in GJ~436 is
shown in Figure 1 along with the same line in four comparison M dwarfs with similar
B-V and V magnitude.  We measure emission at both the \caii\ lines
and find an average value from 2000 to 2004 of $S_{HK} = 0.726$ for
GJ~436 (Wright et al. 2004) on the Mount Wilson scale (Baliunas \etal 1995).
The apparent strong emission in Figure 1 is deceptively striking because
of the weak UV continua of low temperature dwarfs.  Indeed, the ratio
of \caii\ flux to the bolometric flux of the star is only \lrphk ~=~ -5.22
representative of the most chromospherically inactive stars.  However,
the precise values of \rphk ~for M dwarfs remain difficult to measure
and carry uncertainties of $\sim$ 20\% due to the poor calibration of
UV continuum fluxes as a function of B-V for such low mass stars
(Wright et al. 2004).

In Figure 1 we show the {\ion{Ca}{2} K} emission from GJ~436 and
four comparison stars with similar B-V.  For dwarf stars of spectral
type $\sim$M2.5, a chromospheric Mt. Wilson $S$ value of 0.5--1.3 is
typical, showing that emission from the chromospheric gas at
$\sim$10,000 K competes easily with the faint UV continuum of these
cool dwarfs.  The comparison stars have chromospheric emission
bracketing that of GJ~436, rendering them useful comparison stars 
for estimating photospheric jitter and velocity errors.  The velocity
scatter of the comparison stars range from 2.32 and 4.65 \ms over the
past 4 to 6 years, as shown in Figure 2.  A total of 32 program
stars are fainter then $V$ $=$ $10$ with B-V between 1.4 and 1.6
and have at least 10 observations spanning 4 years.  The median
velocity RMS of these 32 stars is 6.7 \mse, including GJ~436 and 
other as yet unknown planet bearing stars.  The four comparison
stars shown in Figures 1 \& 2, along with the other M1.5--M3 dwarfs
of modest chromospheric activity on our planet--search program, show that the
{\it combined} velocity jitter and velocity errors for these
middle-aged M1.5 -- M3 dwarfs having V $>$ 10, is $\sim$ 5 \mse.

\section{Doppler--Shift Measurements}

\subsection{Stellar Sample and Doppler Technique}

We have been monitoring the radial velocities of 150 M dwarfs at the
Keck 1 telescope for the past 4 years.  Most were drawn from the
Hipparcos catalog (ESA 1997), with supporting stellar information
taken from Reid, Hawley \& Gizis (1995).  The sample comprises a
complete sample of isolated M dwarfs (separation greater than 2 arcsec
from any companion) accessible to Keck within 9 pc that are brighter
than V = 11.  The magnitude threshold favors selection of early--type
M dwarfs, M0--M5, causing exclusion of dwarfs later than M5.  A few of
the 150 M dwarfs are fainter than V = 11 and a few are farther than 9
pc.  Our complete sample of M dwarfs has been monitored for the past
3 -- 7 years and is listed by Wright et al. (2004).  For M dwarfs having
magnitudes, V = 8 -- 12, the typical exposure times are 4--8 minutes
yielding a S/N ratio per pixel in the spectra of 300--75,
respectively.  The resulting radial velocity measurements have an
internal precision of 2--8 \ms, based on the agreement (uncertainty in
the mean) of the $\sim$ 400 spectral intervals of 2 \AA.

We measure Doppler shifts by placing an Iodine absorption cell (Marcy
\& Butler 1992) near the focal plane of the telescope centered on the
optical axis, to superimpose iodine lines on the stellar spectrum,
providing a wavelength calibration and proxy for the point spread
function (PSF) of the spectrometer (Valenti et al. 1995).  The
temperature of the cell is controlled to $50.0 \pm 0.1$ C and the
pyrex Iodine cell is sealed so that the column density of iodine
remains constant (Butler \etal 1996).  The Keck Iodine cell was not
altered during the entire duration of the project, preserving the
zero--point of the velocity measurements despite any changes to the
optics of the HIRES spectrometer.  The HIRES spectrometer is operated
with a resolution R $\approx 70000$ and wavelength range of 3700 -- 6200
\AA\ (Vogt et al. 1994) though only the region 4950 -- 6000 \AA\ (with
iodine lines) was used in the Doppler analysis.  The Doppler shifts
from the spectra are determined with the spectral synthesis technique
described by Butler \etal (1996).

Representative sets of velocity measurements for 4 stars with similar
B--V and V magnitude are shown in Figure 2.  These four stars bracket
GJ~436 in both apparent brightness and chromospheric activity.  The
four comparison stars exhibit RMS velocity ranging from 2.32 to 4.65 \mse,
representing the bottom line in the error budget, including errors
from limited photons, any instrumental effects, Doppler analysis
errors, and astrophysical jitter effects for such stars.

\subsection{Velocities of GJ~436}

We obtained 42 high resolution spectra of GJ~436 at the Keck 1
telescope with the HIRES echelle spectrometer (Vogt et al. 1994)
during the 4.5--year period, Jan 2000 to July 2004 (JD = 2451552.1 --
2453196.8).  The times of observation, velocities, and uncertainties
are listed in Table 1.  The exposure times were 8 -- 10 minutes,
yielding S/N~$\approx$~150 and resulting in an uncertainty in the
radial velocity of 4.4 \ms (median) per exposure.  Starting on 29 July
2003 (JD = 2452849), we noticed an apparent periodicity of 2.64 d in
our extant velocities.  During the ensuing year of observations, as we
tested the existence of the prospective planet, we usually obtained
three consecutive exposures within a night to reduce the
photon--limited errors by $\sqrt{3}$ to $\sim$ 3 \mse.  We suspect
that inadequacies in our current deconvolution algorithm result in
another 2 \ms of stochastic error caused by accentuated noise in our
deconvolved spectrum.  Refinement of our deconvolution algorithm for M
dwarfs is in progress.

Figure 3 shows the measured velocities vs. time for GJ~436, with each
point representing the binned velocities in intervals of 2 hours for
clarity.  The internal velocity uncertainty was typically 4.4 \mse
(median) as gauged from the uncertainty in the mean of the 400
spectral chunks separately analyzed in the Doppler analysis.  This
Doppler uncertainty is similar to that of the comparison M dwarfs of
similar V magnitude (Figure 2).  The recently obtained multiple
exposures which were binned to final velocity measurements from 2003.9
to the present, have uncertainties of only $\sim$ 3--4 \mse,
benefitting from the greater number of photons collected.

The velocities for GJ~436 exhibit an RMS of 13.3 \mse.  This scatter
is much greater than the internal errors of 4.7 \ms that stem from the
uncertainty in the mean of the 400 spectral chunks that are separately
analyzed for their Doppler shifts.  The expected jitter is only 3.3
\ms based on comparable M dwarfs, as discussed in section 2.2.  One
may compute the probability that the scatter would be as large or
larger than 13.3 \ms due to chance fluctuations of the known doppler
errors and jitter, added in quadrature.  We adopt here the quadrature
sum of internal Doppler error for each measurement and the expected
jitter of 3.3 \mse, as the effective noise.  We fit a straight line to
the velocities and examine the reduced $\chi^2$ for the residuals,
adopting this effective noise as the uncertainty per measurement in
the calculation of $\chi^2$.  The resulting reduced
$\sqrt{\chi^2_{\nu}}$ = 2.57 which has a probability of occurrence by
chance of less than 0.001.  Thus the velocity scatter in GJ~436 is
larger than can be understood by known sources of errors and
photospheric jitter.

\section{Orbital Analysis}

A periodogram of the entire set of Keck velocities for GJ~436 is
shown in Figure 4.  A strong peak resides at a period of 2.643 d.  The
the false alarm probability associated with this peak is FAP $<$ $10^{-3}$ based on
both the analytical assessment of the number of independent
frequencies (Gilliland and Baliunas 1987) and by Monte Carlo
realizations in which the velocities are scrambled and power spectra
recomputed at the scrambled velocities.  The window function produces
peaks that surround the peak at 2.64 d, but they are not statistically
significant.

We have fit the velocities for GJ~436 with a Keplerian model
including a floating linear velocity trend, as shown in Figure 5.  The
fit yields an orbital period, $P$ = 2.6441 d, velocity semi-amplitude
$K = 18.1 \pm 1.2$ \mse, and an eccentricity of $e$~=~0.12 .  All
orbital parameters are listed in Table 2.  Adopting the stellar mass
of 0.41 \msun (\S 2.1) implies a minimum mass for the orbiting
companion of \msini = 0.067 \mjup and a semi-major axis of 0.0278 AU.
The linear velocity trend has slope of 2.7$\pm$1.5 \ms per year, implying the
possible existence of a more distant companion, but still only
marginally credible.

The uncertainties in Table 2 are based on Monte Carlo realizations of
the data.  The best--fit radial velocity curve is subtracted from the
original velocities and the residuals are adopted as representative of
the amplitude and distribution of velocity noise from all sources,
including velocity errors and photospheric jitter.  The residuals are
permuted and added back to the best-fit radial velocity curve, leaving
the times of observation the same.  This approach yields many
realizations of the set of velocity measurements of the best--fit
planet, assuming that the noise distribution is as exhibited by the
residuals.  A Gaussian distribution of errors was not assumed (but
such a simplification yields similar values for the uncertainties in
the orbital parameters).  Each realization was fit with a Keplerian
model, allowing calculation of the standard deviation of each
orbital parameter.  These are adopted as the 1--$\sigma$
uncertainties, as listed in Table 2.   Note that these quoted
uncertainties do not incorporate the uncertainty in the massof the star itself.

The Keplerian fit yields residuals with a standard deviation of 5.26
\mse, consistent with the expected errors and the RMS for the
comparison M dwarfs (Figure 2).  Similarly, the fit yields
$\sqrt{\chi^2_{\nu}}$ = 1.00, indicating that the Keplerian model from
a single planet is adequate to explain the velocities.  The
eccentricity of 0.12 $\pm$ 0.06 is nearly consistent with a circular
orbit.  Tidal coupling is expected for such a close planet, with an
especially short circularization time scale if the planet is partially
solid.

We attempted fits to the velocities using the simplest model, notably
one with an assumed circular orbit and no allowed velocity trend.
Such models have only three free parameters and the best--fit circular
orbit model is shown in Figure 6.  The resulting best--fit circular
orbit has $P$~=~2.644 d, $K$~=~14.0 \mse, and \msini = 0.052 \mjupe,
implying a mass slightly smaller than that found in the
eccentric--orbit model with a trend.  The circular orbit model (no
trend) yielded residuals with RMS = 6.8 \ms and $\sqrt{\chi^2_{\nu}}$
= 1.25, both somewhat larger than those found from the eccentric
model, but not a large enough difference to rule it out.  We then fit
the velocities with a Keplerian having non--zero eccentricity, but no
trend, which yielded \chisq = 1.20 and $e$ = 0.11.  Finally, we fit
the velocities with a circular orbit, but leaving the trend floating,
which yielded RMS = 6.19 \ms and \chisq = 1.15 , and gave \msini =
0.052 \mjupe.  This model consisting of a circular orbit with floating
trend reduces \chisq \ to a level that is near enough that achieved
with the eccentricity allowed to float that we cannot rule out a
circular orbit.  Thus, a circular orbit remains plausible and implies
a lower planet mass, \msini = 16.5 \mearthe.

\section{False Alarm Probability}

The Keplerian fit to the velocities yields an acceptable value of
\chisq ~=~ 1.03 when including a velocity trend, and yields nearly as
acceptable a value, \chisq ~=~ 1.23, when carrying out a fit with only a
circular orbit and no trend.  Nonetheless, one might be concerned that
the plethora of possible orbits of short periods, less than $\sim$10
d, might permit random fluctuations to yield such low values of
$\chi^2$ by chance.

We tested the hypothesis that the velocities are merely uncorrelated
noise such that the Keplerian fit yields a low $\chi^2$ due merely to
fluctuations of that noise. We have carried out two tests of this null
hypothesis, one using the F--statistic and the other with Monte Carlo
simulations of scrambled velocities.  The F--Test is described by Ford
(2004), Cumming (2004) and Marcy et al. (2004b).  The improvement in
$\chi^2$ between a model that assumes no planet and one that includes
a Keplerian orbit, $\Delta \chi^2$, can be assessed for the
probability that such improvement would occur by chance fluctuations.
We form the ratio $\Delta \chi^2/\chi^2_{\nu}$ which follows the F
distribution (Bevington \& Robinson 2002, Cumming 2004) and permits
assessment of the probability that this ratio departs from 0.0 due to
fluctuations alone.  That probability corresponds to the false alarm
probability (FAP) for the best--fit Keplerian model.  Each independent
frequency (1 / orbital period) can harbor such fluctuations.  We
therefore determine the number of independent frequencies (periods) by
constructing an interval between them such that a phase difference of
one full cycle accrues during the entire time series (Cumming 2004).
This F--Test is essentially identical to the computation of FAP from a
periodogram analysis.  In our test, however, a Keplerian model, rather
than a sinusoid, is compared to the no--planet model.  We applied this
test to GJ~436.  We find that the probability that \chisq\ improved
due to mere fluctuations of noise from 2.57 (no planet) to 1.03
(Keplerian plus trend) is less than 1$\times$10$^{-5}$.  We thus find
it unlikely that noise fluctuations can account for the low \chisq\
found from the Keplerian model.
 
The F--Test cannot properly account for the non--uniform sampling of
the velocities nor the non--Gaussian nature of the velocity errors.
Therefore, we have carried out another test of FAP that involves
scrambling the velocities, as if they were uncorrelated noise, and
recomputing a Keplerian fit for each scrambled realization of the
data.  In this way, we determine the distribution of $\chi^2$ that is
expected if the velocities were simply uncorrelated noise.

We scrambled the velocities, keeping the times of observation the
same.  For each of 1000 realizations, we searched for the best--fit
Keplerian model and recorded its associated value of $\chi^2$.  The
resulting histogram of $\sqrt{\chi^2_{\nu}}$ is shown in Figure 7 and
shows the distribution expected if the measured velocities were simply
uncorrelated noise. The distribution peaks at $\sqrt{\chi^2_{\nu}}$ =
2.05 with a width, FWHM = 0.3.  None of the 1000 trials of scrambled
velocities yielded a value of $\sqrt{\chi^2_{\nu}}$ below the
best--fit $\sqrt{\chi^2_{\nu}}$ of 1.03 from the original velocities.
Thus, if the measured velocities are simply noise, the probability is
$<$1/1000 that the value of $\chi^2_{\nu}$ for the best--fit Keplerian
orbit planet would be caused by chance fluctuations.  Moreover the
distribution of $\sqrt{\chi^2_{\nu}}$ from scrambled velocities is so
well separated from the value of $\sqrt{\chi^2_{\nu}}$ from the
original velocities (Figure 7), that the FAP is likely to be significantly 
less than 0.001, consistent with the even lower FAP value found from the
F--Test.  This robust Monte Carlo analysis suggests directly that
fluctuations in uncorrelated noise cannot account for the low $\chi^2$
from the Keplerian fit.  With a stellar sample of 150 M dwarfs on the
Keck planet search, the probability is low that such chance
fluctuations might arise in any one of the M dwarfs.  Thus, the
velocities appear inconsistent with the hypothesis that the high
quality of the Keplerian fit stems merely from chance fluctuations.

We also considered that systematic errors might account for the velocity
variations seen in GJ~436.  We have computed periodograms from our
velocity measurements of the other 150 M dwarfs obtained with HIRES
during the past four years.  None shows a periodicity anywhere near a
period of 2.6 d with any amplitude close to the 18 \ms seen here.
Thus, we see no evidence of any instrumental source of a 2.6--day
periodicity.  For the same reason, no other M dwarfs reveal any
evidence of intrinsic {\em astrophysical} periodicities at that
period.  Similarly, the other 1100 FGK stars on our Keck
planet--search program show no evidence of 2.6--day periodicities.
Indeed, the shortest period planet found on this program is that of HD
46375 with a period of 3.02 d, rendering the 2.6--day period of GJ~436
clearly extraordinary in our Doppler planet survey.

One conceivable source of periodicity is the rotation of the star that
could modify the spectrum due to any inhomogeneities on the stellar
surface.  However, we see no photometric periodicity in GJ~436 at
periods near 2.64 d (Figure 8) at millimag levels.  This lack of
brightness variations suggests the absence of large spots and active
regions distributed non--uniformly over the photosphere, with limits on
the covering factor of under 1\% .  The star clearly has a
chromosphere, as seen in the \caii\ emission (Figure 1), which is
likely distributed in patches over magnetic regions.  However, the
Doppler information in optical spectra comes from the photosphere
which apparently has uniform surface brightness, with fluctuations no
more than 1\% judging from the constant photometry.

If the 2.644--day Doppler period were the rotation period of star, the
implied equatorial velocity would be 7.3 \kms (by adopting a stellar
radius of 0.38 R$_{\odot}$) (Chabrier \& Baraffe 2000).  Such a large
equatorial velocity could be detectable in the rotational broadening
of the absorption lines.  We compared the widths of the lines in
GJ~436 to those in a comparison star, GJ 411 which has
V$\sin i <$ 2 \kms (Chen and Marcy 1992).  We find that the lines
in GJ~436 are no broader than those in GJ 411, giving an upper
limit of 3 \kms on V$\sin i$.  Thus the hypothesis that GJ~436
is rotating at nearly 7.3 \kms appears to be unlikely.  Stellar
rotation seems unlikely to be the cause of the Doppler periodicity
at $P$ = 2.644 d.

\section{Photometric Observations}

Queloz et al. (2001) and Paulson et al. (2004) have shown that photospheric features 
such as spots and plages on solar-type stars can result in low-amplitude, 
periodic radial velocity variations capable of mimicing the presence of a 
planetary companion.  Therefore, precision photometric measurements are an 
important complement to Doppler observations and can help to establish 
whether the radial velocity variations are caused by stellar magnetic 
activity or planetary-reflex motion, e.g., Henry et al. (2000a).  Photometric 
observations can also detect possible transits of the planetary companions 
and so allow the determination of their radii and true masses, e.g.,
Henry et al.(2000b).

We have observed GJ~436 with the T12 0.8~m automatic photometric telescope 
(APT) at Fairborn Observatory between 2003 November and 2004 June and 
obtained a total of 226 brightness measurements.  The T12 APT is equipped 
with a two-channel precision photometer employing two EMI 9124QB bi-alkali 
photomultiplier tubes to make simultaneous measurements in the Str\"omgren 
$b$ and $y$ passbands.  The APT measures the difference in brightness between 
a program star and a nearby constant comparison star with a typical precision 
of 0.0015 mag for bright stars ($V < 8.0$).  For GJ~436, we used the 
comparison star HD~102555 ($V$ = 7.24, $B-V$ = 0.39, F2), which was shown to 
be constant to 0.002 mag or better by comparison with the second comparison 
star HD~103676 ($V$ = 6.79, $B-V$ = 0.38, F2).  We reduced our Str\"omgren 
$b$ and $y$ differential magnitudes with nightly extinction coefficients and 
transformed them to the Str\"omgren system with yearly mean transformation 
coefficients.  Further information on the telescope, photometer, observing 
procedures, and data reduction techniques employed with the T12 APT can be 
found in Henry (1999) and in Eaton, Henry, \& Fekel (2003).

The 226 combined $(b+y)/2$ differential magnitudes of GJ~436 are
plotted in the top panel of Figure~8.  The observations are phased
with the planetary orbital period and a time of inferior conjunction,
computed from the orbital elements in Table~2.  The standard deviation
of the observations from the mean brightness level is 0.0043 mag,
larger than the typical 0.0015 mag precision with the T12 APT, because
GJ~436, at $V$ = 10.67, is much fainter than the typical star observed
with this telescope.  By averaging the Str\"omgren $b$ and $y$
observations into a single passband, we gained a factor of square root
2 in our precision, improving our sensitivity to any intrinsic stellar
variability.  Period analysis does not reveal any periodicity between
1 and 100 days.  A least-squares sine fit of the observations phased
to the radial velocity period gives a semi-amplitude of 0.00044 $\pm$
0.00037 mag.  Thus starspots are unlikely to be the cause of the
velocity periodicity.  If the star were pulsating with a velocity
amplitude of 18 \mse, the difference between its minimum and maximum
radius would be 1300 km, yielding a fractional change in disk size of
0.005.  Thus, the observed very low limit to possible photometric
variability supports planetary-reflex motion as the cause of the
radial velocity variations.  Note that even if the planet were as
large as Jupiter, it would intercept only 3$\times$10$^{-4}$ of the
star's radiation.

The observations near phase 0.0 are replotted with an expanded abscissa in 
the bottom panel of Figure~8.  The solid curve in each of the two panels 
approximates the predicted transit light curve assuming a planetary orbital 
inclination of 90\arcdeg ~(central transits).  The out-of-transit light level 
corresponds to the mean brightness of the observations.  The transit duration 
is calculated from the orbital elements.  Four different transit depths are 
estimated from an assumed stellar radius of 0.41 R$_{\sun}$, a planetary 
mass of 1.2 M$_{NEP}$, and planetary radii of 0.5, 0.35, 0.31, and 
0.24 R$_{JUP}$, corresponding to planetary models of a gas giant without a 
core, a gas giant with a core, an ice/rock planet, and a planet composed of 
pure iron, respectively.  The horizontal bar below the predicted transit 
window in the bottom panel represents the approximate uncertainty in the 
time of mid transit, based on Monte Carlo simulations and the uncertainties 
in the orbital elements.  The vertical error bar to the right of the transit 
window corresponds to the $\pm$ 0.0043 mag measurement uncertainties for a 
single observation.  The geometric probablility of transits is 6.8\%, computed 
from equation 1 of Seagroves et al. (2003) assuming random orbital inclinations.  
The mean of the 8 observations within the transit window agrees with the 
mean of the 218 observations outside the window to within 0.0010 mag, just 
as expected from the precision of the observations.  Thus, central transits 
for the four planetary models given above would produce transit depths of 
17, 8, 7, and 4 sigma, respectively.  Although the uncertainty in the time 
of mid transit is somewhat larger than the duration of possible transits, 
the observations nonetheless rule out the possibility of complete (as
opposed to grazing) transits except possibly for shallow events occuring 
around phase 0.99 for a planet with a rocky or iron composition.  Since the
planet lies at a distance of 14 stellar radii, the inclination of the
orbit must be less than about 86\arcdeg.

\section{Discussion}

The radial velocities of GJ~436 exhibit a marked periodicity,
consistent with a Keplerian orbit of a planetary-mass companion.  No
other interpretation, such as stellar oscillations or rotational
modulation of surface inhomogeneities, seems likely to explain the 2.6
d periodicity.  The implied Keplerian orbit has a period of 2.644 d,
an orbital semimajor axis of 0.0278 AU, and an orbital eccentricity of
0.12 that is marginally consistent with circular.

The stellar mass of 0.41 \msun implies a minimum planet mass, \msini,
of 0.067 \mjup or 1.2 \mnep .  This minimum mass is considerably lower
than that of any extrasolar planet previously found (pulsar planets
aside).  The lowest previously found planet, as of this writing, had been that of
HD 49674 with \msini = 0.11 \mjup (Butler et al. 2003).  For randomly
oriented orbits, the average value of $\sin i$ is $\pi/4$ and it is
probable that $\sin i > 0.5$.  Thus it is likely that this planet has
a mass less than 2 \mnepe.  

A planet of roughly Neptune mass orbiting 0.028 AU from an M dwarf
raises several new issues about its constitution.  We could not rule
out the possibility of a solid rock or rock--ice composition, nor a
primarily ice--rock body with a significant hydrogen envolope
reminiscent of Neptune and Uranus in our Solar System.  Indeed, one
wonders if a gaseous envelope can be ruled out for this planet on the
basis of its survival against UV energy deposition from the young,
magnetic M dwarf.  The uncertainty in its composition leaves a range
of plausible radii for the planet, from 0.2 -- 1.0 \rjupe (especially
for arbitrary orbital inclinations), leaving
uncertain the amount of dimming expected by
transits.  The planet intercepts 3 $\times$ 10$^{-4}$ of the star's
radiation, if it has the radius of Jupiter.  This would be the
amplitude of reflected light variations if the planet's albedo were
unity and the orbit were edge--on.

After submission of this paper, another Neptune-mass planet (\msini =
0.82 \mnep) emerged from Doppler measurements made by the Hobby Eberly
Telescope and Lick Observatory (McArthur et al. 2004).  Orbiting
55 Cancri (G8~V), this other neptune raises similar questions about
its origin, migration, and composition.  The existence of two planets
having \msini near the mass of Neptune drastically reduces the already
remote statistical possibility that face--on orbital inclinations
explain the low values of \msini for them .  Instead, it is likely
that a population of Neptune--mass planets exists that is the
extension of the rising mass function already known toward lower
planetary masses.

From its \mv \ of 10.63 and expected bolometric correction of --1.9,
the luminosity of GJ~436 is $L$~=~0.025 \lsun . At its orbital distance
of 0.028 AU, the expected surface temperature is $\sim$ 620 K,
depending on its albedo, greenhouse effects, and uniform illumination,
all of which are questionable.  Most metals and refractory material
remain solid at such a temperature which is similar to that on the
surface of Venus.  Insignificant mass loss would occur from the tail
of the Maxwell--Boltzmann distribution of hydrogen.  However, a
detailed calculation is required to determine the mass loss from the
planet due to high--energy stellar radiation from flares and the
corona, especially during first billion years of enhanced magnetic
activity on the star.  The possibility of Roche--lobe overflow (in
either direction) especially during pre--main--sequence evolution,
should also be considered.  Tidal coupling must be computed to
determine if the planet keeps one hemisphere toward the star.  If the
planet were mostly solid, questions would be raised about the
temperature on both the back side and on the terminator.

This star is just the second M dwarf known to harbor a planet, the
first being two--planet system around GJ 876 (Marcy et al. 2001).  At
a distance of 10.2 pc, GJ~436 is a prime target for the Space
Interferometry Mission (SIM) to detect the astrometric wobble and
place limits on $\sin i$ and hence the planet mass. Coronagraphic
imaging missions from the ground and space should attempt to image
planets residing farther from this star, especially because of the
velocity trend that seems to be preferred in our model.

We find that giant planets are rare among M dwarfs.  Among the 150 M
dwarfs in our Keck survey, this star is only the second found to have
a planet despite three years of surveying them with high Doppler precision of 3
\mse.  The low mass of this new planet highlights our ability to detect
planets of somewhat higher mass, 0.3 \mjup or greater within 1 AU
($P ~<~ $1.5 yr) during which time at least two orbits would have
transpired.  However, only one M dwarf has revealed such a
jupiter--sized planet around an M dwarf.  Thus, the occurrence rate of
Jupiter--mass planets within 1 AU of M dwarfs appears to be
1/150~$\approx$~0.7\%.  In contrast, among our 1180 FGK stars surveyed
at the Lick, Keck, and AAT telescopes, 41 have a planet within 1
AU. Thus, for nearby FGK stars, the occurrence rate of jupiter--mass
planets (0.5~$<~M~<$~13 \mjup) within 1 AU is 3.5\% (Marcy et
al. 2004a).  Thus, the occurrence of jupiters orbiting near M dwarfs
appears to be a factor of $\sim$~5 below that of solar--mass stars.

This paucity of giant planets around M dwarfs is consistent with, but
not required by, the planet--formation models of Levison, Lissauer, \&
Duncan (1998) and of Alibert, Mordasini, \& Benz (2004).
Lower mass protoplanetary disks around M dwarfs may have been
important in slowing the accretion rate, yielding lower--mass planets
(Laughlin et al. 2004).  Indeed, the formation of Neptune in our Solar
System is not well understood and may have been influenced by the low
surface mass density in the outer Solar nebula (Lissauer et al. 1995;
Bryden, Lin, \& Ida 2000; Thommes, Duncan, \& Levison 2002).  Thus, it
appears that the occurrence of jupiters is a function of stellar mass.
Stars more massive than the Sun may harbor Jupiter-mass planets
in greater numbers and masses than found so far around Solar--type
stars.

\acknowledgements 
We thank John Johnson, and Chris McCarthy for help
with the observations and analysis, and we thank Eugene Chiang and
Peter Bodenheimer for valuable conversations.  We gratefully
acknowledge the superb dedication and support of the Keck Observatory
staff. We appreciate support by NASA grant NAG5-75005 and by NSF grant
AST-0307493 (to SSV); support by NSF grant AST-9988087, by NASA grant
NAG5-12182 and travel support from the Carnegie Institution of
Washington (to RPB). GWH acknowledges support from NASA grant NCC5-511
and NSF grant HRD-9706268. JJL acknowledges support from NASA's Solar
Systems Origins grant 188-07-1L.  We are also grateful for support by
Sun Microsystems. We thank NASA and the University of California for
allocations of Keck telescope time toward the planet search around M
dwarfs.  This research has made use of the Simbad database, operated
at CDS, Strasbourg, France. Finally, the authors wish to extend thanks
to those of Hawaiian ancestry on whose sacred mountain of Mauna Kea we
are privileged to be guests. Without their generous hospitality, the
Keck observations presented herein would not have been possible.

{}

\clearpage

\begin{figure}[t!]
\epsscale{0.85}
\plotone{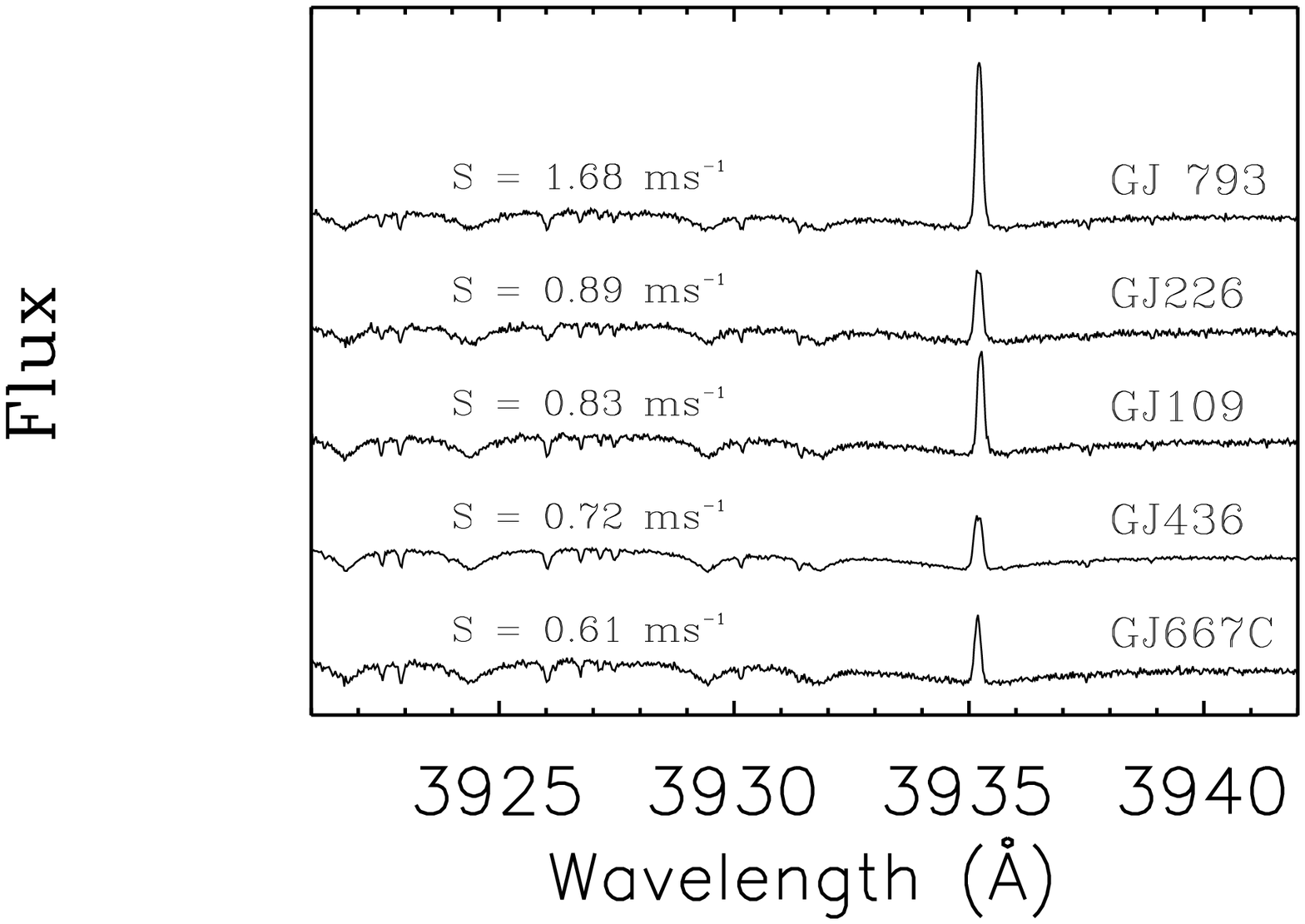}
\caption{Spectra of the chromospheric Ca II K emission line for GJ~436 
and four comparison M dwarfs with the similar B-V and V magnitude.
The stars are plotted in the ascending order of chromospheric
S value: GJ~667C ({\em bottom}), GJ~436, GJ~109, GJ~226,
and GJ~793 ({\em top}).  These four comparison stars have chromospheric
emission that bracket that of GJ~436, rendering them good comparison stars
for GJ~436.}
\label{fig1}
\end{figure}
\clearpage

\begin{figure}
\plotone{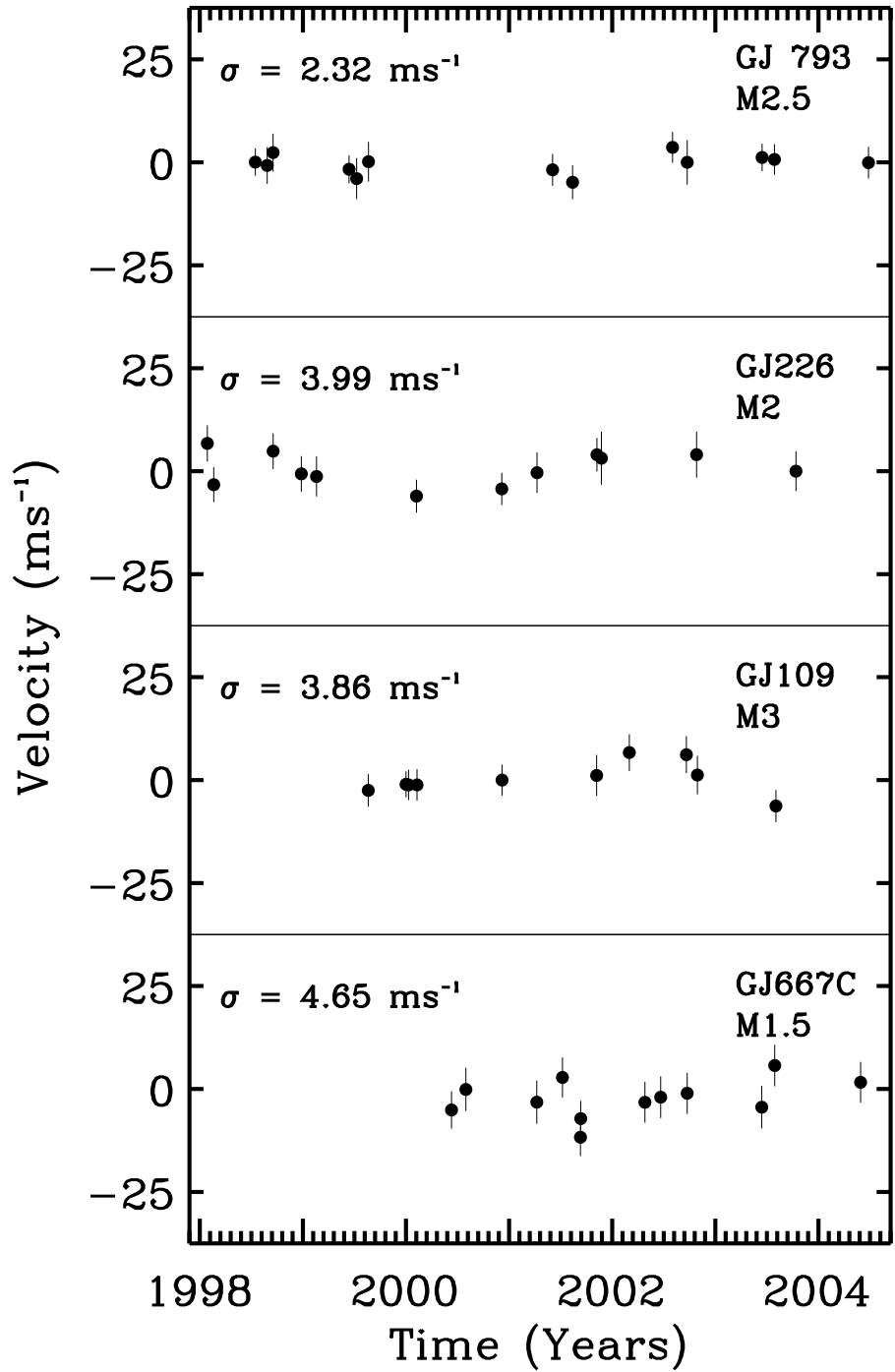}
\caption{Radial velocities vs. time of the comparison M dwarfs 
shown in Figure 1.  These M dwarfs are representative of the 
middle--aged M1.5 -- M3 dwarfs on the program.  These stars have
10+ observations over 4+ years.  The observed RMS velocity
scatter of these stars range from 2.32 to 4.65 \mse, showing the
combined velocity errors and photospheric jitter is $\sim$5 \ms
or less, suggesting that GJ~436 will suffer similar errors.}
\label{fig2}
\end{figure}
\clearpage

\begin{figure}
\plotone{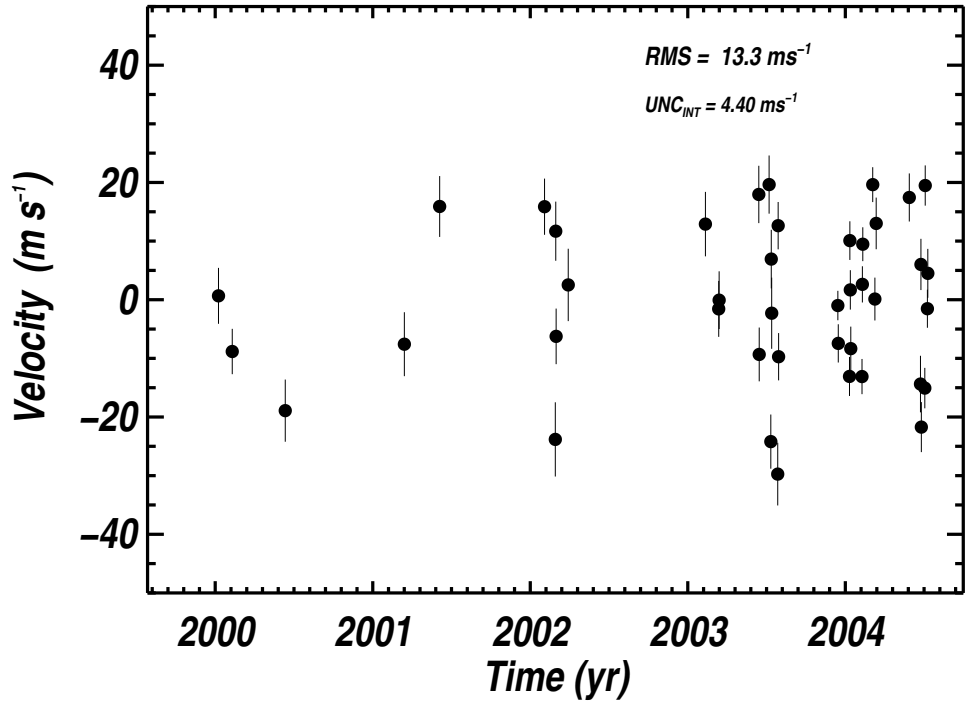}
\caption{Radial velocities vs. time for GJ~436.  The observed RMS
velocity scatter of 13.3 \ms is larger than both the median of the
internal errors, 4.7 \mse, and the expected RMS ( 5 \ms) revealed by
comparison stars (Figure 2).  Similarly, the expected photospheric
jitter is only 3.3 \ms.  The value of $\sqrt{\chi^2_{\nu}}$ = 2.57, for
which the probability of occurrence by chance is less than 0.1\%.
Thus the velocity scatter in GJ~436 is larger than can be
understood by sources of errors and jitter.  The velocity zero--point
is arbitrary.}
\label{fig3}
\end{figure}
\clearpage

\begin{figure}
\plotone{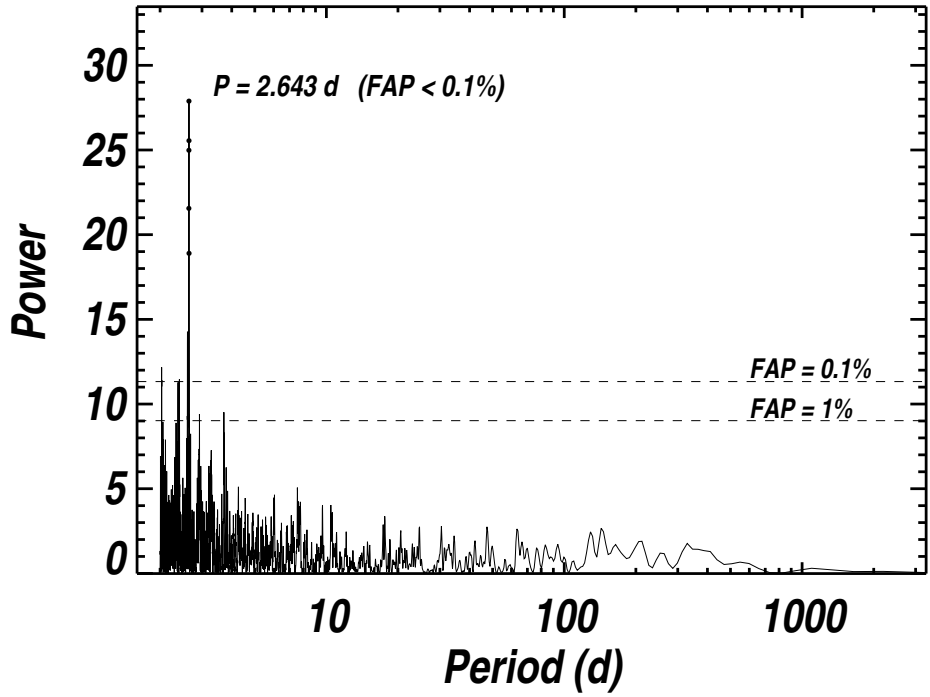}
\caption{The periodogram of the Keck velocities for GJ~436,
showing peak power at 2.643 d with a false alarm probability, FAP $<$
0.1\%. The multiple dots near the highest peak show the
sampling that resolves the peak.  The neighboring peaks are
aliases of the 2.64 d period.}
\label{fig4}
\end{figure}
\clearpage

\begin{figure}
\plotone{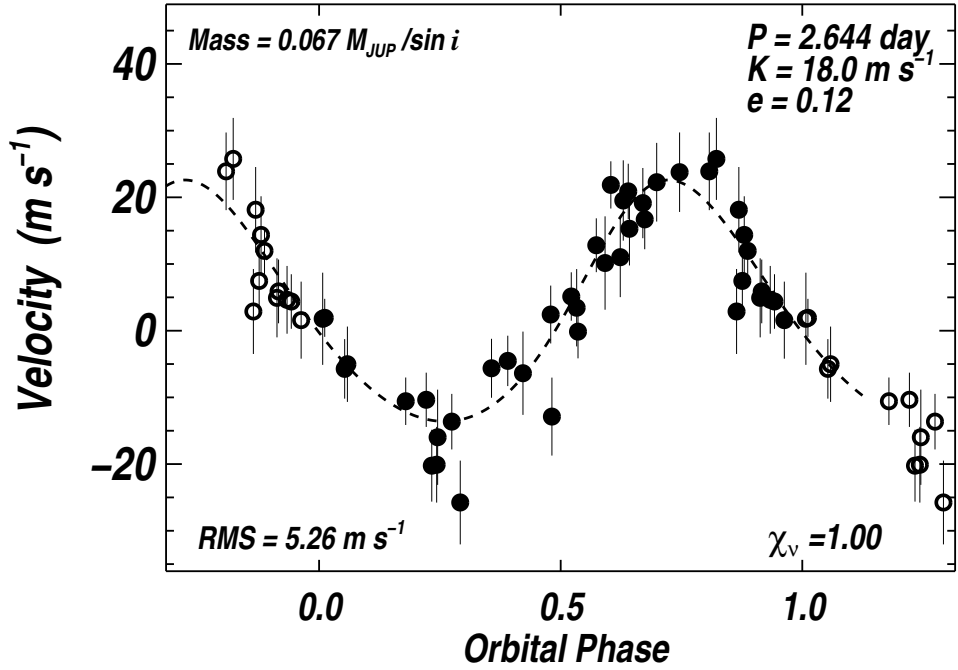}
\caption{Measured velocities vs. orbital phase for  GJ~436 (filled dots), with
repeated points (outside phases 0--1) shown as open circles.
The dotted line is the radial velocity curve from the best--fit
orbital solution, $P$ = 2.644 d, $e$ = 0.12, \msini = 0.067 \mjupe. The
RMS of the residuals to this fit is 5.26 \ms with a reduced
$\sqrt{\chi^2_{\nu}}$ = 1.00. The error bars show the quadrature sum
of the internal errors (median 5.2 \mse) and jitter (3.3 \mse).  A
linear velocity trend is found to be 2.7 \ms per year.}
\label{fig5}
\end{figure}
\clearpage

\begin{figure}
\plotone{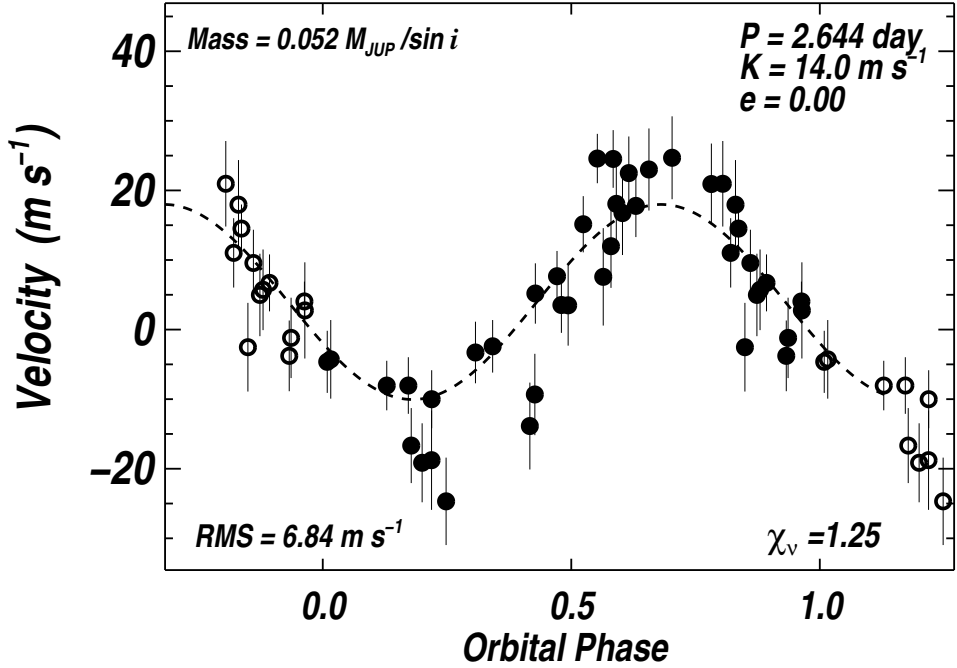}
\caption{Circular orbit fit to GJ~436, overplotted on the measured
  velocities vs. orbital phase (dots, as in Figure 5).  
  The dotted line represents the
  sinusoidal fit (circular orbit) and no linear velocity trend,
  allowing only three free parameters.  This orbital fit gives $P$ =
  2.644 d, $e$ = 0.0 (forced), $K$=14.0 \mse, \msini = 0.052 \mjupe. The
  RMS of the residuals is 6.8 \ms with a reduced $\sqrt{\chi^2_{\nu}}$
  = 1.25, indicating a somewhat poorer fit than for a full Keplerian
  fit with a floating eccentricity and linear velocity trend (Fig
  5). The weights and error bars reflect the quadrature sum of the
  internal errors (median 5.2 \mse) and jitter (3.3 \mse).}
\label{fig6}
\end{figure}

\clearpage
\begin{figure}
\plotone{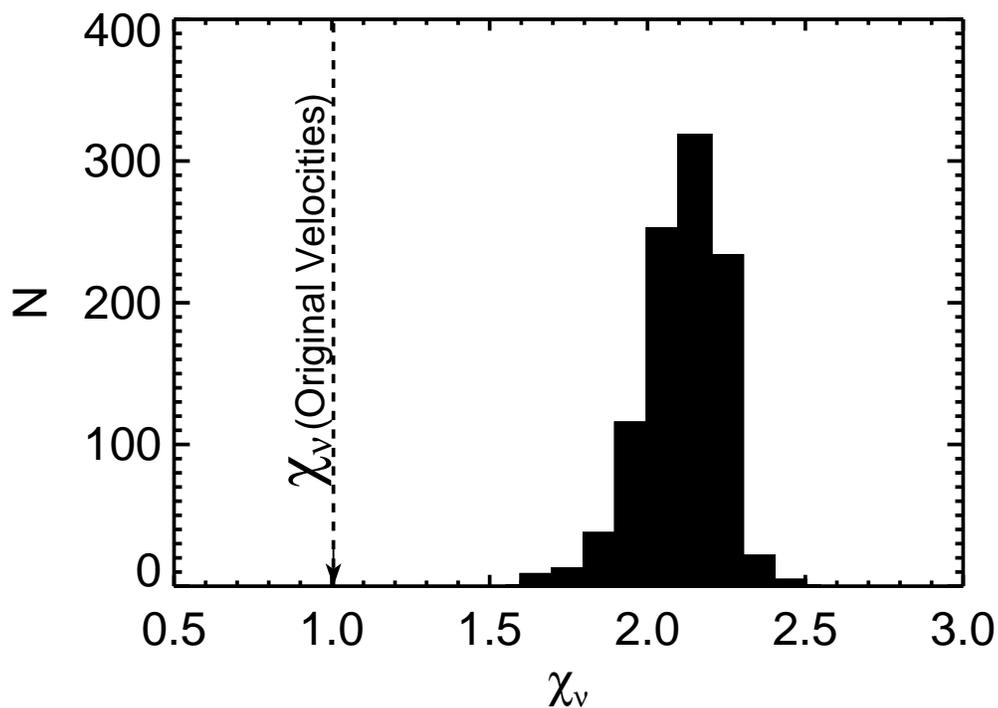}
\caption{Histogram of $\sqrt{\chi^2_{\nu}}$ from Keplerian fits to
1000 sets of scrambled velocities (filled area).  The histogram shows
the probability distribution of $\chi_{\nu}$ from Keplerian fits that
would occur if the velocities were merely uncorrelated noise.  The
distribution peaks at $\chi_{\nu}$ = 2.1 with a FWHM of 0.3.  For
comparison, the best--fit orbit to the original velocities gives
$\sqrt{\chi^2_{\nu}}$ = 1.00 (vertical dashed line), which is lower
than all 1000 trials.  The FAP is apparently much less than 0.1\%, in
agreement with the F--test that yields FAP $<$ 0.1\%.}
\label{fig7}
\end{figure}

\clearpage
\begin{figure}[t!]
\figurenum{8}
\epsscale{0.85}
\plotone{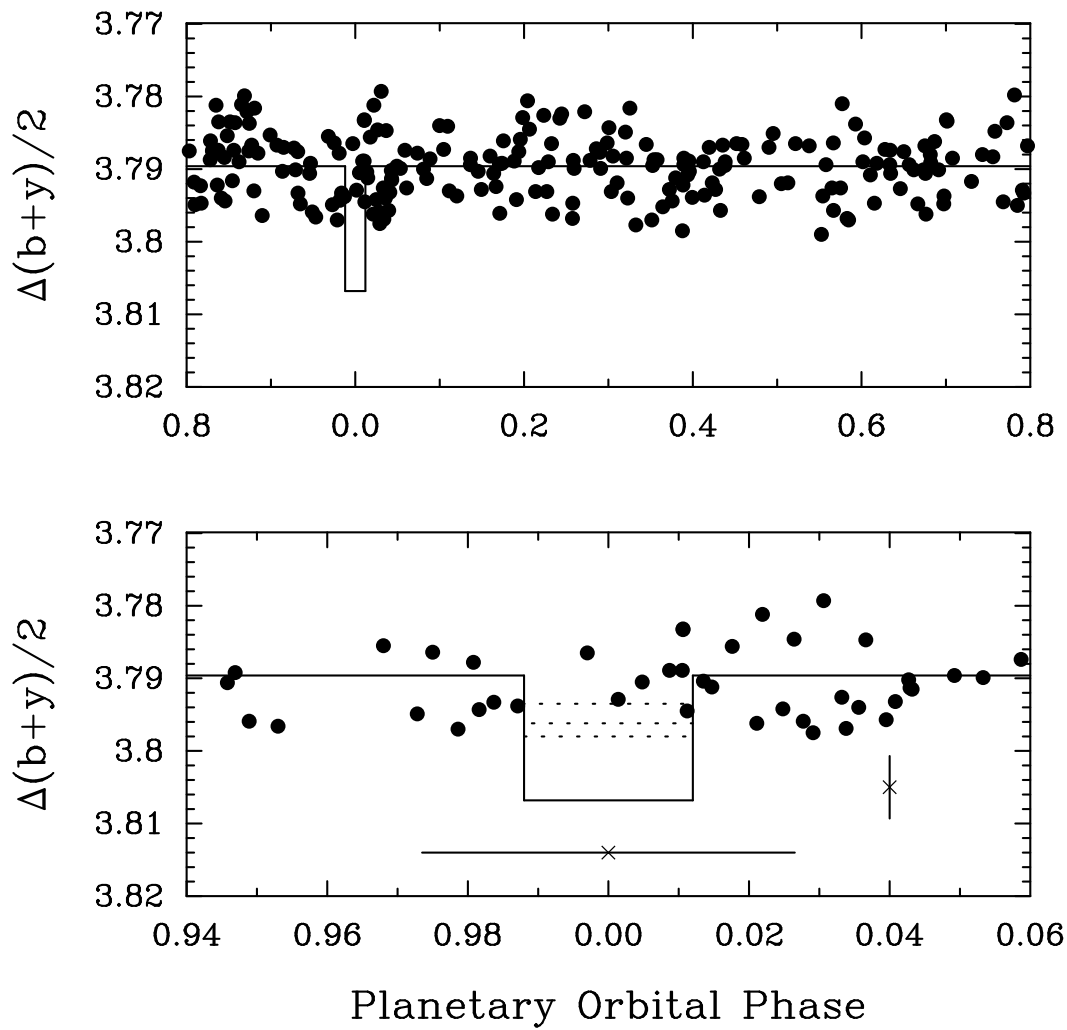}
\figcaption{Str\"omgren $(b+y)/2$ photometric observations of GJ~436
acquired with the T12 0.8~m APT at Fairborn Observatory phased to the
Doppler perioditiy of 2.644 days ($top$).  
In particular, the star is constant on the radial velocity period to a 
limit of 0.0004 mag or better, supporting the planetary interpretation of
the radial velocity variations.  Predicted transit depths are shown
($bottom$) for various planetary compositions (see text), but are ruled out
by the observations.}
\end{figure}

\clearpage

\begin{deluxetable}{rrr}
\tablenum{1}
\tablecaption{Radial Velocities for GJ~436}
\label{}
\tablewidth{0pt}
\tablehead{
\colhead{JD}         & \colhead{RV}     & \colhead{Unc.}  \\
\colhead{-2440000}   & \colhead{(\ms)}  & \colhead{(\ms)} 
}
\startdata
 11552.077 &    0.66 &    4.8  \\
 11583.948 &   -8.83 &    3.9  \\
 11706.865 &  -18.91 &    5.3  \\
 11983.015 &   -7.59 &    5.5  \\
 12064.871 &   15.90 &    5.2  \\
 12308.084 &   15.87 &    4.8  \\
 12333.038 &  -23.82 &    6.3  \\
 12334.054 &   11.69 &    5.0  \\
 12334.935 &   -6.25 &    4.7  \\
 12363.039 &    2.53 &    6.2  \\
 12681.057 &   12.90 &    5.5  \\
 12711.898 &   -1.56 &    4.8  \\
 12712.902 &   -0.07 &    4.9  \\
 12804.878 &   17.96 &    4.9  \\
 12805.829 &   -9.33 &    4.6  \\
 12828.800 &   19.64 &    5.0  \\
 12832.758 &  -24.20 &    4.6  \\
 12833.763 &    6.93 &    5.0  \\
 12834.779 &   -2.30 &    6.1  \\
 12848.752 &  -29.74 &    5.3  \\
 12849.762 &   12.63 &    4.0  \\
 12850.763 &   -9.72 &    4.0  \\
 12988.146 &   -0.99 &    2.5  \\
 12989.146 &   -7.45 &    3.3  \\
 13015.142 &  -13.08 &    3.3  \\
 13016.072 &   10.08 &    3.3  \\
 13017.046 &    1.68 &    3.4  \\
 13018.142 &   -8.34 &    3.8  \\
 13044.113 &  -13.10 &    3.0  \\
 13045.018 &    2.62 &    3.1  \\
 13045.984 &    9.47 &    2.9  \\
 13069.032 &   19.63 &    3.0  \\
 13073.992 &    0.12 &    3.7  \\
 13077.066 &   13.02 &    4.4  \\
 13153.817 &   17.44 &    4.1  \\
 13179.759 &  -14.38 &    4.8  \\
 13180.803 &    6.02 &    4.4  \\
 13181.746 &  -21.72 &    4.3  \\
 13189.787 &  -15.07 &    3.4  \\
 13190.754 &   19.48 &    3.4  \\
 13195.767 &   -1.53 &    3.3  \\
 13196.772 &    4.50 &    4.2  \\
\enddata
\end{deluxetable}
\clearpage

\begin{deluxetable}{lr}
\tablenum{2}
\tablecaption{Orbital Parameters for GJ~436}
\label{param}
\tablewidth{0pt}
\tablehead{
\colhead{Parameter} & \colhead{} \\
}
\startdata
 $P$ (d)         		   &  2.6441 (0.0005)        \\
${\rm T}_{\rm p}$ (JD)     &  2451551.507 (0.03)  \\
$e$                          &  0.12 (0.06)         \\
$\omega$ (deg) 	           &  332 (11)            \\
$K_1$ (\ms)    		   &  18.10 (1.2)           \\
$f_1$(m) (M$_\odot$)       &  1.58e-12             \\
\arel (AU)                 &  0.0278                \\
$M\sin i$ (M$_{Jup}$)      &  0.067 (0.007)         \\
{\rm d}$v/{\rm d}t$ (\ms per yr)  &  2.7             \\
${\rm Nobs}$               &  42                   \\
RMS (\ms)                  &  5.26                  \\
$\sqrt{\chi^2_{\nu}}$      &  1.00                 \\
\enddata
\end{deluxetable}
\clearpage
\end{document}